# Information Encoding with Optical Dielectric Metasurface via Independent Multichannels

*Fengliang Dong, [†,‡] Hang Feng, [§,‡] Lihua Xu, [†,‡] Bo Wang, [§] Zhiwei Song, [†] Xianfeng Zhang, [†] Lanqin Yan [†] Xiaojun Li,[†] Lianfeng Sun,\*, [†] Yan Li, \*, [§, ‖] and Weiguo Chu\*, [†]*

[†] CAS Key Laboratory of Nanosystem and Hierarchical Fabrication, Nanofabrication Laboratory, CAS Center for Excellence in Nanoscience, National Center for Nanoscience and Technology, Beijing 100190, China.

[§]State Key Laboratory for Mesoscopic Physics, Department of Physics, Peking University and Collaborative Innovation Center of Quantum Matter, Beijing 100871, China.

[‖]Collaborative Innovation Center of Extreme Optics, Shanxi University, Taiyuan, Shanxi 030006, China

KEYWORDS: *dielectric metasurface*, *multiwavelength*, *photon spin-dependent state*, *Pancharatnam-Berry phase, multichannel*, *information processing*

ABSTRACT: Information encryption and security is a prerequisite for information technology which can be realized by optical metasurface owing to its arbitrary manipulation over the wavelength, polarization, phase and amplitude of light. So far information encoding can be implemented by the metasurface in one dimensional (1D) mode (either wavelength or polarization) only with several combinations of independent channels. Here we successfully apply dielectric metasurfaces in a 2D mode (both wavelength and polarization) with far more combinations of independent channels to encrypt information, which therefore enhances the encryption security dramatically. Six independent channels by two circular polarization states (RCP and LCP) and three visible wavelengths (633 nm, 532 nm and 473 nm) in 2D mode can produce 63 combinations available to information encoding, in sharp contrast with 7 combinations by 3 independent channels in 1D mode. This 2D mode encoding strategy paves a novel pathway for escalating the security level of information in multichannel information encryption, anti-counterfeiting, optical data storage, and information processing.

Optical metasurfaces, consisting of plasmonic or dielectric meta-atoms or meta-molecules on a substrate[1-7], can manipulate the wavefront of incident light through an abrupt phase change with subwavelength resolution. The arbitrary controllability over the amplitude, phase and polarization of light, and the easy fabrication and compact configuration endow metasurfaces with great potential and strong competitiveness in development of miniaturized, exotic and integrated optical components. The state of the art fabrication techniques have enabled numerous metasurface-based planar optical components with diversified functionalities in the infrared and visible wavelengths to be realized such as arbitrary reflection and refraction[1, 8, 9], focusing of light[6, 7, 10-19], hologram[4, 5, 20-26], nonlinear dynamics[27], strong photonic spin Hall effect[3, 4, 28], invisible cloak carpet[29], and polarization control and analysis[30-33].

With the well recognized importance of information security, many cryptography techniques based on classical and quantum optics have been developed[34, 35]. Optical cryptography is considered to have great potential in information security because of the availability of four degrees of freedom, such as wavelength, polarization, amplitude and phase, which can secure information more reliably by encrypting with various combinations. Recently, metasurfaces have been reported for image encoding with only one unitary element of either polarization or wavelength (one dimensional mode, 1D). For example, ultrathin nonlinear plasmonic metasurfaces have been used to encrypt images at a near infrared wavelength[36, 37]. The encoded information can be detected by the visible light with a specific wavelength[36] or the broadband wavelengths[37] under the illumination of the coherent near infrared light due to second harmonic generation wave. In addition, we have also evidenced the capability of the metasurfaces encoding information without crosstalk by encoding color images into silicon dielectric metasurfaces composed of subwavelength metamolecules formed by independently multiplexing three nanoblocks corresponding to red, green and blue, respectively in which three primary light fields are well confined within the corresponding nanoblocks and manipulated individually.[23] Image encoding based on polarization-controlled metasurfaces involves photon-spin channels (right and left circular polarization: RCP and LCP)[4, 38] or a pair of orthogonal linear polarization states[5, 7]. Here, a couple of channels are formed by two orthogonal polarizations (Say, RCP / LCP for circular, and 0º / 90º for linear), and the information encoded in one is independent of and undetectable by the other. Recently, we have adopted this concept to realize the color-tunable holograms.[39] Compared to 1D mode, information encryption in 2D mode using both polarization and wavelength increases the flexibility and capacity of encryption, which would make the encoded information more secure. Therefore, information encryption in 2D mode is expected to have potential applications in information security.

Here we design reflective-type silicon metasurfaces in the visible region to encode information into independent channels using both polarization and wavelength. The metasurfaces consist of elliptical nanocylinder meta-atoms with three sizes which correspond to three different wavelengths and are independently multiplexed in a subwavelength supercell. The encrypted information is converted into phase profiles using the Gerchberg-Saxton algorithm and then encoded into the orientation angle of the meta-atoms on holographic metasurfaces by use of Pancharatnam–Berry (PB) phase. We experimentally demonstrate that the information encoded in a specific channel, for example, the wavelength of 633 nm and RCP, is read only by RCP red light whereas the information is undetectable by either LCP red, green or blue light or for RCP green or blue. Moreover, the information can be encoded with various combinations of polarization (RCP, LCP) and visible wavelengths (red, green and blue). Six independent channels with the wavelengths of 633 nm, 532 nm and 473 nm, and the orthogonal polarization states of RCP and LCP allow one to derive 63 combinations in total which can be applied to information encoding.

**Results and Discussion**

**Design of metasurfaces.** Figure 1 shows the design of metasurfaces encoding independently the characters of 'NCNST' and 'PKU' in 2D mode using both wavelength and polarization. We choose the wavelengths of red (633 nm), green (532 nm) and blue (473 nm), and the pair of orthogonal photon-spin states (RCP and LCP) as 6 independent channels to encode information. The characters 'NCNST' are encoded to be read by RCP light of three wavelengths, 633 nm ('NT'), 532 nm ('CS'), and 473 nm ('N'), respectively, and the characters 'PKU' by red ('P'), green ('K'), and blue ('U') LCP light, respectively. Therefore, when illuminating the metasurface with RCP light of 633 nm, 532 nm, and 473nm, respectively, red 'NT', green 'CS', and blue 'N' can be read independently (Figure 1a). In contrast, red 'P', green 'K', and blue 'U' are reconstructed with LCP light at the corresponding wavelengths (Figure 1b). Figures 1c shows the geometry of a holographic metasurface supercell composed of elliptical nanocylinders with three sizes in a subwavelength scale lattice constant of $P$=350 nm which is intentionally designed to reduce the higher order diffractions. The lengths of the major and minor axes of meta-atoms ($l_x$, $l_y$) for the wavelengths of 633 nm, 532 nm and 473 nm are 145 and 118 nm, 126 and 90 nm as well as 102 and 80 nm, respectively, and their heights all are $h$=230 nm. One can find that two elliptical nanocylinders for blue light are designed to have the same size of $l_x$=102 nm and $l_y$=80 nm for increasing the diffraction efficiency. In the supercell, elliptical nanocylinders of three sizes are multiplexed independently. We choose the sizes of the nanocylindrical meta-atoms to minimize the possible crosstalk between the wavelengths. [39] The diffraction efficiency, as an important parameter of performance, were simulated for both an individual meta-atom corresponding to each wavelength and the super-cell containing four meta-atoms, as shown in in Figure 1d. It can be seen that though the efficiency for 473 nm, 532 nm and 633 nm wavelength may not achieve the highest (8%, 62%, 30%), respectively, the crosstalk is minimized and negligible due to the low efficiencies of one wavelength – specific meta-atom manipulating the other wavelengths. With the coupling between the meta-atoms considered, i. e. for the supercell, the efficiency of each wavelength changes (5%, 43%, and 33% for 473, 532 and 633 nm, respectively, as indicated by the black line in Figure 1d)), which is however still comparable with that of an individual meta-atom only.

**Independence of information encoding.** As pointed out above, the information encoded in either of a pair of orthogonal photon-spin states (RCP and LCP) is independent of and thus undetectable by the other. Meanwhile, we as well realize the invisibility of the information encoded at one wavelength to others by trimming the sizes of multiplexed elliptical nanocylinders. To demonstrate the independence of the channels in 2D mode, we perform simulations on a periodic array consisting of multiplexed meta-atoms with two sizes to manipulate LCP green and RCP red light, respectively. Each has eight elliptical nanocylinders with a step size of phase retardation of $\pi/4$ to generate full $2\pi$ PB phase (Figure 2a). In the supercell, the period is set to $P$=320 nm with the sizes of elliptical nanocylinders of $l_x$=156 nm, $l_y$=128 nm and $l_x$=118 nm, $l_y$=92 nm corresponding to the wavelengths of 633 nm and 532 nm, respectively. Despite the weak manipulation over the other wavelength which is revealed by the near field distributions (Figure S1) and the propagating field distributions (Figure S2b, c), LCP green and RCP red light can still be taken as the independent manipulation by the array of multiplexed two-size meta-atoms (Figure 2b, c), as the array consisting of single size meta-atoms does (Figure S2a, d). We design and fabricate a metasurface hologram by encoding an image with four channels formed by two colors (red and green) and two polarization states (RCP and LCP), as shown in Figure 2. The red part of the image is designed for RCP red light and the green part is encoded for LCP green light. The diffraction efficiencies of the nanoblocks with the sizes chosen for red and green are simulated to be 22% and 70%, respectively (Figure S3a). Although the selected sizes don't give rise to the maximum diffraction efficiencies, the wavefront crosstalk between two wavelengths is significantly reduced due to the weak coupling between the two-size meta-atoms, as also evidenced by Figure 2. Figures 2g and 2h show SEM images of the fabricated metasurface. We experimentally obtained the red part of the target image for RCP red light and the green part for LCP green light (Figure 2i-k), in accordance with the simulations (Figure 2e, f). The simultaneous illumination of RCP red and LCP green light produces the expected results, which is unambiguously revealed by Figure 2k.

To elucidate the independence of 2D channels, the invisibility of the information encoded into one channel to others is demonstrated. For both RCP and LCP states, we reconstructed the hologram images by scanning the wavelength from 470 nm to 690 nm using a tunable supercontinuum laser source (NKT- SuperK EXU-6), as shown in Figure 3 and Figure S4. Figure 3 clearly shows that the information encoded in the channels of both RCP red (630nm) and LCP green (530nm) is readable only by RCP red (Figure 3c) and LCP green (Figure 3e) light, respectively, whereas the information is invisible to other wavelengths (Figure 3a, 3b and 3d, 3f). Furthermore, we verified experimentally polarization-specific independence of information encoded in both RCP red and LCP green channels. Figure 4 shows the results of information reconstruction at the wavelengths of 633 nm and 532 nm. For the information encoded into the channels of RCP red and LCP green, the intensities of the corresponding reconstructed images can be expressed as $I_r \cos^2 \varphi$ and $I_g \sin^2 \varphi$, respectively, where $I_r$ and $I_g$ are the reflected intensities for RCP red (633 nm) and LCP green (532 nm), respectively, and $\varphi$ is the angle between the axes of the reconstruction light and the RCP light (see Supporting Information). The vanishing intensities in Figure 4e and 4f indicate that the information encoded in a polarization channel is invisible to the orthogonal polarization state.

The crosstalk between different wavelengths for multi-spectral wavefront manipulation by a single metasurface is inevitable, which originates from the intrinsic dispersion of the meta-atom material. Namely, the information encoded at a specific wavelength can be detected by others for the broadband or multispectral response[25, 26]. A broadband of 120 nm for both RCP (550-670 nm) and LCP (490-610 nm) is observed in Figure S4. The information encoded in the RCP red channel can be seen at the wavelength of 490 nm as well.

**Six channels for information encoding.** To increase the flexibility of information encryption, we design and fabricate a metasurface containing 6 independent channels formed by three visible wavelengths and two polarization states, as elucidated in Figure 1. Figures 5a and 5b are top view and tilt view SEM images of the fabricated metasurface with elliptical cross sections, respectively. When the metasurface is illuminated with RCP red, green and blue light simultaneously, only 'NCNST' can be reconstructed (Figure 5c), and the image of 'PKU' can be seen with LCP light at the same three wavelengths (Figure 5d).

With the 6 independent channels built in the 2D mode of wavelength and polarization, we can encrypt the information in a single channel, as shown in Figure 6a-f, or a combination of the independent channels (Figure 6g-j). The total possible combinations $C_{total}$ can be derived as

$$C_{total} = C_6^1 + C_6^2 + \cdots + C_6^6 = 2^6 - 1 = 63 \qquad (1)$$

where $C_n^k$ is a k-combination that has n elements. Undoubtedly, the more the combinations of independent channels are, the more secure the encoded information is. The channel independence of the information encoded allows one to decode and process the information individually and directly using a coherent light with a specific polarization state at a particular wavelength conveniently, as revealed in Figure 6. This would greatly improve the efficiency of information processing and reduce its cost, in sharp contrast with use of a filter as usually done.

The 1D mode with either polarization or wavelength can only provide a few independent channels and thus quite limited combinations for information encryption, whereas for the 2D mode with both polarization and wavelength, more independent channels and then one order of magnitude more combinations can be achieved. Balthasar Mueller et al[38] and Arbabi et al[7] encoded different information into 2 independent channels of RCP /LCP and *x*/*y*, respectively. For 1D mode simply with polarization, a pair of orthogonal polarization states (circular, elliptical, or linear) can form only 3 combinations with 2 independent channels at most for information encryption. Likewise, for 1D mode only with wavelength we encoded color images into 3 channels at 633 nm, 532 nm and 473 nm independently [23] to achieve only 3 independent channels and thus 7 combinations as well. However, the 2D mode with 3 visible wavelengths and 2 circular polarization states has been demonstrated to build the combinations as many as 63 with 6 independent channels, which allows for more flexible and reliable information encryption. This can be generalized to have the combinations as many as $2^N$-1 for $N$ independent channels, which would escalate the security level of information encryption and data encoding dramatically.

One should bear in mind that 2D mode information encoding holds for not only the reflective dielectric metasurfaces on SOI substrates but also the transmissive dielectric metasurface on transparent substrates such as glass, as shown in our previous work [23]. The multiplexibility of meta-atoms in a supercell also makes it possible to extend the wavelength range from visible to infrared to even tetrahertz for information encoding. In addition, pairs of orthogonal linear or elliptical polarization states other than circular polarizations can be adopted as the encoding channels to increase the number of combinations. To think along this line, use of both multiple wavelengths and polarizations as independent channels can undoubtedly boom the number of combinations, which would upgrade the security level of information drastically as well as enlarge the capacity of information processing enormously. Also, we would like to point out that for the PB phase based information coding, 'negative' information with a phase delay of π (a ghost image [26]) would be obtained when the orthogonal polarization light with the same wavelength illuminates the metasurface with information encoded with a specific wavelength and polarization state. The ghost image may be removed by encoding information adopting the combination of propagation phase and geometric (PB) phase. [38, 40] Our purpose is to present an information coding approach with independent multichannels instead of the coding methodology responsible for the ghost image.

In summary, we propose the scheme of 2D mode using both wavelength and polarization as independent channels to boom the number of combinations for information encryption based on photon spin-dependent dielectric metasurfaces, which can escalate the security level of encoded information drastically. The independent information encoding in six channels from three visible wavelengths (633 nm, 532 nm, 473 nm) and two orthogonal circular polarization states (RCP and LCP) is experimentally demonstrated using the metasurface - consisting of silicon elliptical nanocylinder meta-atoms with three different sizes. The realization of independent information encoding results from the independent manipulability and effective confinement of light at three primary wavelengths by the meta-atoms. The multiple combinations of polarization and wavelength in 2D mode capable of greatly increasing the information encoding capacity and dramatically upgrading the security level of information encryption can be applied to not only the reflective but the transmissive metasurfaces. The power and flexibility of this strategy shows the great potential of optical metasurfaces in multichannel information encryption, anti-counterfeiting, optical data storage, and information processing.

**Methods**

**Numerical simulations.** We performed the 3D simulations based on a finite element method using commercial software COMSOL Multiphysics. The near field distributions of electric, magnetic and power fields, reflective field distributions and diffraction efficiencies were characterized using periodic boundary conditions. The refractive index of the 230 nm thick Si film is shown in Figure S3c, and that of 2 μm thick oxide layer is set as 1.46 in the simulation.

**Sample fabrication.** The reflective-type metasurfaces are fabricated on the SOI substrates. First, the positive electron beam resist (ZEP-520A) with a thickness of about 200 nm is spin coated on the sample and baked at 180 ℃ on a hotplate for 2 minutes. Then, elliptical nanostructures are defined on the resist film based on the standard electron beam lithography (EBL, Vistec EBPG 5000+) and the following development process. The metasurface with elliptical nanocylinder meta-atoms is fabricated by etching the 230 nm thick silicon top layer of SOI using a mixture of $SF_6$ and $O_2$ gases on an Inductively Coupled Plasma Etcher (Plasmalab System 100, ICP180). Finally, the residual resist on the top of meta-atoms is removed in an ultrasonic bath of butanone solvent.

**Optical characterization.** The experimental setup for sample characterization is shown in Figure S5. A tunable supercontinuum laser source (NKT- SuperK EXU-6) is employed to provide the light with specific wavelengths in the visible regime from 470 nm to 690 nm (Figure S5a). The beam with RCP or LCP is realized using a combination of a polarizer and a quarter-wave plate which is then focused by a lens with a focal length of $f$=120 mm. The metasurface is located on the focal spot, and the reconstructed images are projected on a screen and captured using a camera. The information encoded in the channels with different polarization states is reconstructed with a setup capable of individually controlling the polarization state (Figure S5b).

ASSOCIATED CONTENT

**Supporting Information**. The Supporting Information is available free of charge on the ACS Publications website at DOI: 10.1021/acsphotonics.XXXXXXX.

Simulation and theory of independent channels; Information encryption in 2D mode; Simulated diffraction efficiencies of different elliptic nanocylinders; Wavelength dependence of encrypted information; Setup of optical characterization. (PDF)

AUTHOR INFORMATION

**Corresponding Author**

*Email: wgchu@nanoctr.cn

*Email: li@pku.edu.cn.

*Email: slf@nanoctr.cn

.

**ORCID**

Fengliang Dong: 0000-0001-6008-7011

Yan Li: 0000-0003-0607-3166

**Author Contributions**

F.D., H.F. and W.C. conceived the idea; F.D. and B.W. performed the designs andsimulations; F.D., L.X., Z.S., and X.L. fabricated the metasurfaces; H.F., B.W. L.Y and X.Z. assisted the measurements and analysis; W.C., Y.L. and L. S. supervised the analysis, experiments, and edited the manuscript. All authors discussed the results and implications and commented on the manuscript at all stages.

‡These authors contributed equally.

**Notes**

The authors declare no competing financial interest.

**Acknowledgments**

This work was supported by Youth Innovation Promotion Association CAS under Grant No. 2015030; CAS Key Technology Talent Program; The Ministry of Science and Technology of China under Grant Nos. 2015DFG62610; The National Natural Science Foundation of China under Grant Nos. 11627803, 11474010 and 61590933; The National Basic Research Program of China under Grant No.2013CB921904 and Key Research Program of Frontier Sciences, CAS (No. QYZDB-SSW-SYS031).

(38) Balthasar Mueller, J. P.; Rubin, N. A.; Devlin, R. C.; Groever, B.; Capasso, F., Metasurface Polarization Optics: Independent Phase Control of Arbitrary Orthogonal States of Polarization. *Phys Rev Lett* **2017**, *118* (11), 113901.

(39) Wang, B.; Dong, F.; Yang, D.; Song, Z. ;Xu, L.; Chu, W.; Gong, Q.; Li, Y., Polarization-controlled color-tunable holograms with dielectric metasurfaces. *Optica* **2017**, *4*(11), 1368-1371.

(40) Wang, B.; Dong, F.; Feng, H.; Yang, D.; Song, Z. ;Xu, L.; Chu, W.; Gong, Q.; Li, Y., Rochon-Prism-Like Planar Circularly Polarized Beam Splitters Based on Dielectric Metasurfaces. *ACS Photonics* **2018**, *5*, 1660-1664.19

Figure captions:

**Figure 1** Schematic diagram of a metasurface for information encryption in 2D mode in the visible region. Red 'NT', green 'CS', and blue 'N' are encoded for RCP, and red 'P', green 'K', and blue 'U' for LCP. The metasurface builds 6 independent channels to form 63 combinations for encoding information, greatly increasing the flexibility and security of encryption. (**a**) Reconstructing 'NCNST' with RCP red, green, and blue illumination, respectively. (**b**) Reconstructing 'PKU' with LCP red, green, and blue illumination, respectively. (**c**) Schematic tilt and top views of elliptic nanocylindrical meta-atoms forming a super-cell. $P$=350 nm, $h$=230 nm, and the sizes ($l_x$, $l_y$) of meta-atoms for the wavelengths of red, green, and blue are (145 nm, 118 nm), (126 nm, 90 nm), and (102 nm, 80 nm), respectively. (**d**) Diffraction efficiencies of multiplexing three-size meta-atoms ($P$=350 nm). The red, green and blue lines correspond to the efficiencies of the above three meta-atoms individually set in a super-cell, and the black line shows the efficiency of the multiplexed meta-atoms in a super-cell.

**Figure 2.** Hologram encoded with the two wavelengths of red and green, and the polarization states of RCP and LCP. (**a**) A meta-molecule of a periodic array consisting of multiplexed meta-atoms with two sizes manipulating the LCP green and RCP red light, respectively. The dashed square denotes a supercell of $P$=320 nm. (**b**) Reflective electric field distribution with LCP green illumination. (**c**) Reflective electric field distribution with RCP red illumination. (**d**) A target image. (**e**) Simulation of red part of the encoded image for RCP. (**f**) Simulation of green part of the encoded image for LCP. (**g**) SEM image of the fabricated hologram. (**h**) SEM image in oblique view. The scale bars are 500 nm in both (**g**) and (**h**). (**i**) Experimental result using RCP illumination at the wavelength

of 633 nm corresponding to (**e**). (**j**) Experimental result using LCP illumination at 532 nm corresponding to (**f**). (**k**) Reconstructed result with the simultaneous illumination of RCP red and LCP green light.

**Figure 3.** Wavelength independence of information encrypted in the fabricated metasurface. The information is encoded in the RCP red and LCP green channels. (**a** - **c**) Experimental results only with the RCP state at the wavelengths of 470 nm, 530 nm, and 630 nm, respectively. (**d** - **f**) Experimental results only with the LCP state at the corresponding wavelengths, respectively.

**Figure 4.** Polarization independence of information encrypted in the fabricated metasurface. The information is encoded in the RCP red and LCP green channels. (**a** - **e**) Reconstructed images illuminating at $\lambda$=633 nm changing the polarization state from RCP to LCP. (**f** - **j**) Reconstructed images at $\lambda$=532 nm with the polarization state from RCP to LCP.

**Figure 5.** Hologram encoded with the three wavelengths of red, green and blue, and the polarization states of RCP and LCP. (**a**) SEM image of the fabricated hologram. (**b**) SEM image in oblique view. The marked meta-atoms in a supercell in red, green and blue correspond to those responsible for the manipulation of red, green and blue light, respectively. The scale bars are 500 nm in both (**a**) and (**b**). (**c**) Reading 'NCNST' by the simultaneous illumination of RCP red, green, and blue light. (**d**) Reading 'PKU' by the simultaneous illumination of LCP red, green, and blue light.

**Figure 6.** Reconstructing the information encoded into different independent channels or their combinations using both wavelength and polarization in 2D mode. (**a** - **c**) 'NT',

'CS', and 'N' in the channel of RCP red, green, and blue, respectively. (**d** - **f**) 'P', 'K', and 'U' in the channel of LCP red, green, and blue, respectively. (**g**) 'NCST' in the combination of RCP red and green. (**h**) 'CNS' in the combination of RCP green and blue. (**i**) 'PU' in the combination of LCP red and blue. (**j**) 'NTK' in the combination of RCP red and LCP green.

Figure 1

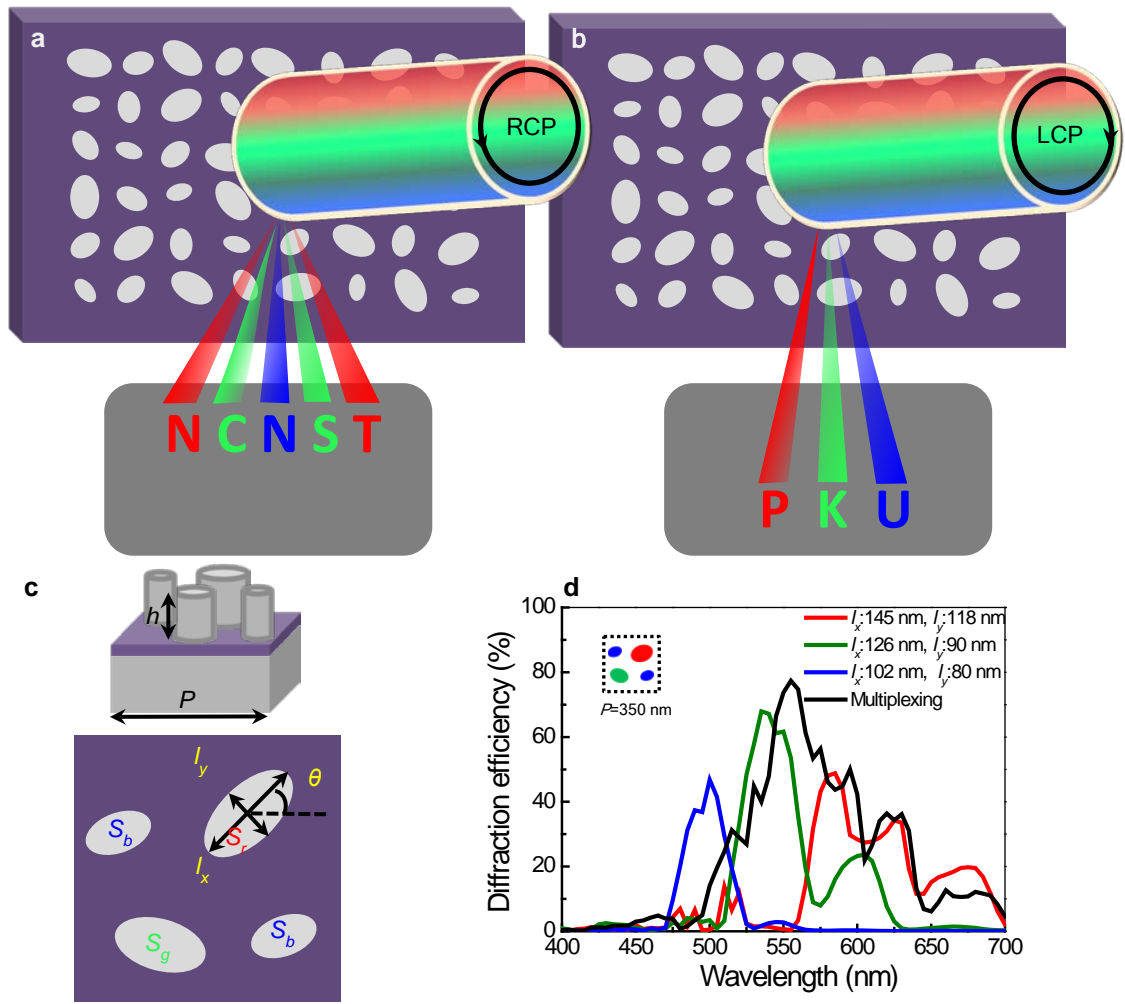

Figure 2

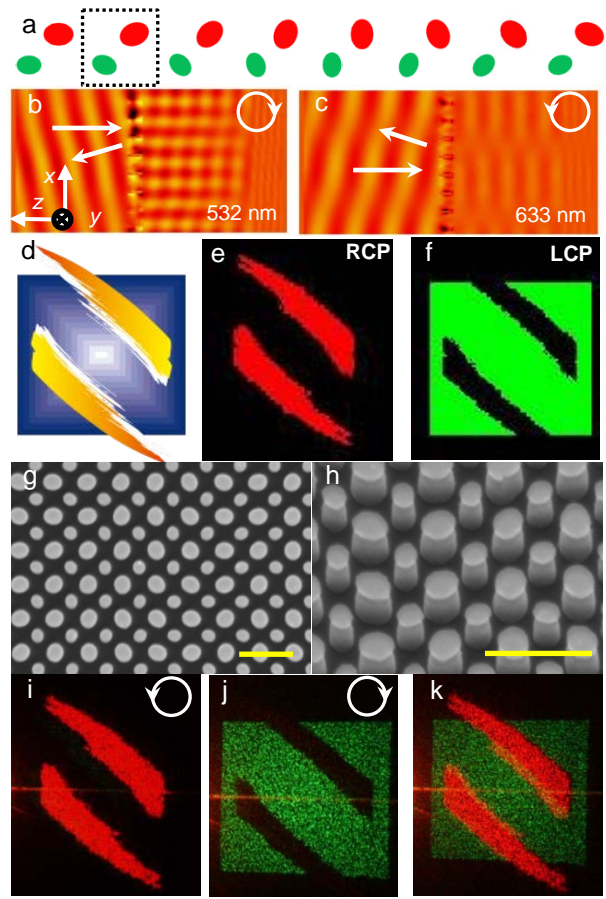

Figure 3

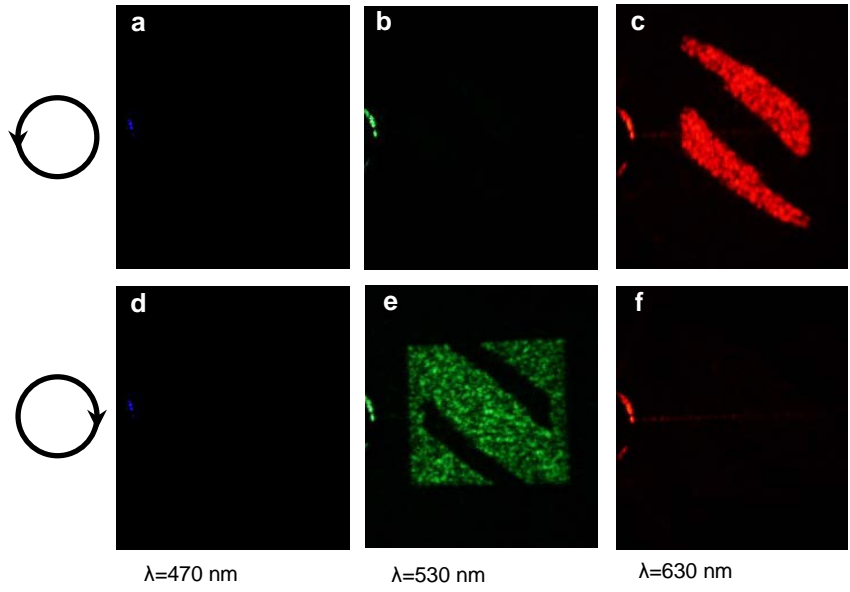

λ=470 nm   λ=530 nm   λ=630 nm

Figure 4

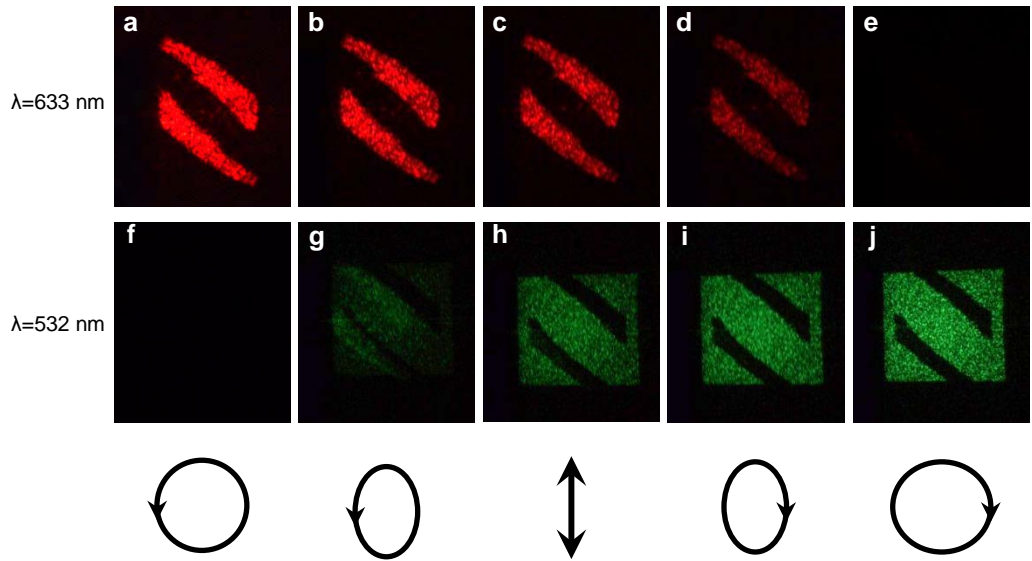

Figure 5

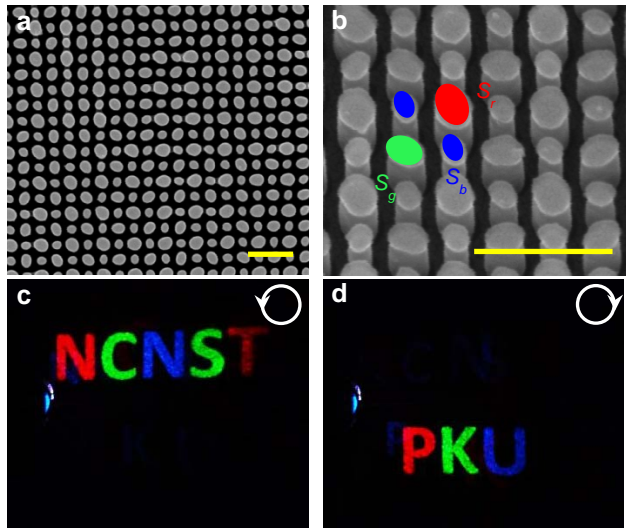

Figure 6

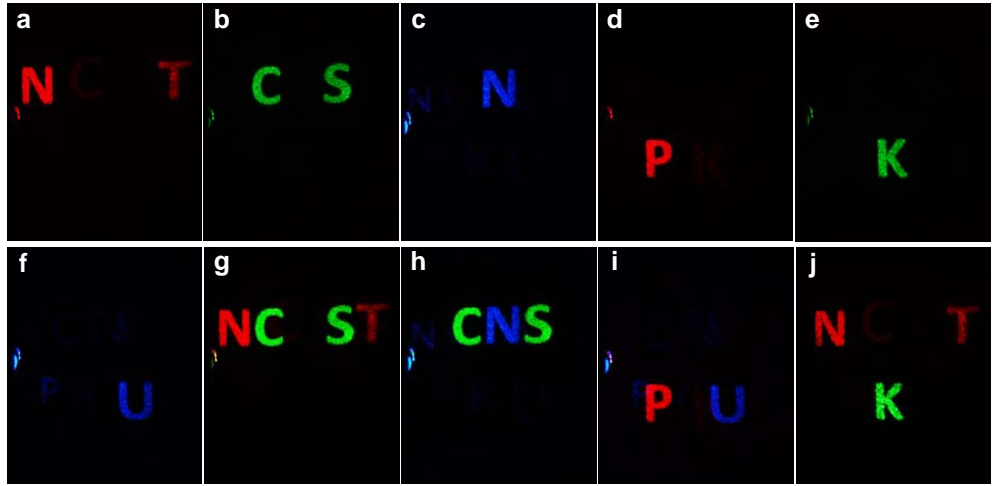

For Table of Contents Only

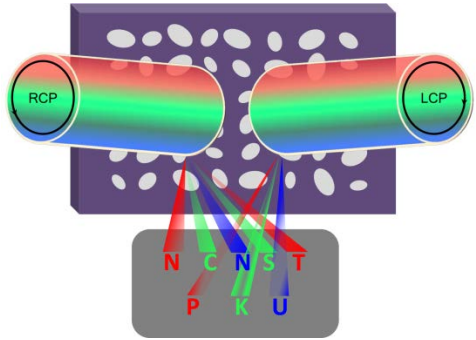

Supporting Information for

# Information Encoding with Optical Dielectric Metasurface via Independent Multichannels

*Fengliang Dong, †,‡ Hang Feng, §, ‡ Lihua Xu, †,‡ Bo Wang, § Zhiwei Song, † Xianfeng Zhang, † Lanqin Yan † Xiaojun Li,† L. F. Sun, \*, † Yan Li, \*,§, ‖ and Weiguo Chu\*, †*

† Key Laboratory of Nanosystem and Hierarchical Fabrication, Nanofabrication Laboratory, CAS Center for Excellence in Nanoscience, National Center for Nanoscience and Technology, Beijing 100190, China.

§State Key Laboratory for Mesoscopic Physics, Department of Physics, Peking University and Collaborative Innovation Center of Quantum Matter, Beijing 100871, China.

‖Collaborative Innovation Center of Extreme Optics, Shanxi University, Taiyuan, Shanxi 030006, China

* Email: wgchu@nanoctr.cn

* Email: li@pku.edu.cn

* Email: slf@nanoctr.cn

**1. Simulation and theory of independent channels**

The simulated near field distributions of electric, magnetic and power fields of multiplexed two-size meta-atoms are shown in Figure S1. With the incident light at the wavelength of 532 nm, the meta-atom with $l_x$=156 nm and $l_y$=128 nm trims the light slightly, and does not affect the manipulation of light by the other meta-atom, just like the effect of the meta-atom of $l_x$=118 nm and $l_y$=92 nm on the manipulation of the incident light at $\lambda$=633 nm.

We compare electric field distributions of the periodic array consisting of multiplexing two-size and single size meta-atoms (Figure S2). One should bear in mind that the Pancharatnam-Berry geometric phase manipulates the circularly polarized light same as the incident light for reflective-type metasurfaces, which is independent of wavelength. However, the manipulation at a specific wavelength depends on the diffraction efficiency. As shown in Figure S2a, d, the manipulation of the arrays of meta-atoms ($l_x$=118 nm and $l_y$=92 nm) for LCP green and meta-atoms ($l_x$=156 nm and $l_y$=128 nm) for RCP red is dominant, respectively. The trimming of the wavefronts of LCP green and RCP red light by the array of multiplexing two-size meta-atoms is equivalent to that by single size meta-atoms individually, evidencing that two color channels are independent of each other.

With the information encoded in a specific polarization state (polarizer) using the illumination with the intensity of $I_0$, the intensity of $I$ in a different polarization (analyzer) follows the Malus' Law:

$$I = I_0 \cos^2 \alpha , \qquad (1)$$

where α is the angle between the axes of the polarizer and the analyzer. For the information encoded in RCP, the intensity will be zero with LCP illumination (α=90°), implying that a pair of orthogonal polarization states build two independent channels.

**2. Information encryption in 2D mode with both wavelength and polarization (3 wavelengths and 2 polarization states)**

It has been pointed out that six independent channels are built to encode information in the 2D mode with 3 wavelengths (473 nm, 532 nm, 633 nm) and 2 orthogonal polarization states (RCP, LCP). The channels are denoted as $T_{Rr}$, $T_{Rg}$, $T_{Rb}$, $T_{Lr}$, $T_{Lg}$, and $T_{Lb}$ where the first subscript stands for the polarization state, RCP or LCP, and the second subscript represents the wavelength of red, green or blue. We can code information in a single channel, or combined channels. There are 6 cases available for coding as follows.

Case 1: a single channel 6 combinations, $C_6^1$, i. e. $T_{Rr}$, $T_{Rg}$, $T_{Rb}$, $T_{Lr}$, $T_{Lg}$, and $T_{Lb}$.

Case 2: two channels

15 combinations, $C_6^2$, i. e. $T_{Rr}+T_{Rg}$, $T_{Rr}+T_{Rb}$, $T_{Rr}+T_{Lr}$, $T_{Rr}+T_{Lg}$, $T_{Rr}+T_{Lb}$, $T_{Rg}+T_{Rb}$, $T_{Rg}+T_{Lr}$, $T_{Rg}+T_{Lg}$, $T_{Rg}+T_{Lb}$, $T_{Rb}+T_{Lr}$, $T_{Rb}+T_{Lg}$, $T_{Rb}+T_{Lb}$, $T_{Lr}+T_{Lg}$, $T_{Lr}+T_{Lb}$, and $T_{Lg}+T_{Lb}$.

Similarly, case 3: three channels with 20 combinations, i.e. $C_6^3$.

Case 4: four channels with 15 combinations, i.e. $C_6^4$.

Case 5: five channels with 6 combinations, i.e. $C_6^5$.

Case 6: six channels with only one possibility, i.e. $C_6^6$.

Then, the combinations in total are

$$C_{total} = C_6^1 + C_6^2 + \cdots + C_6^6 = 2^6 - 1 = 63.$$

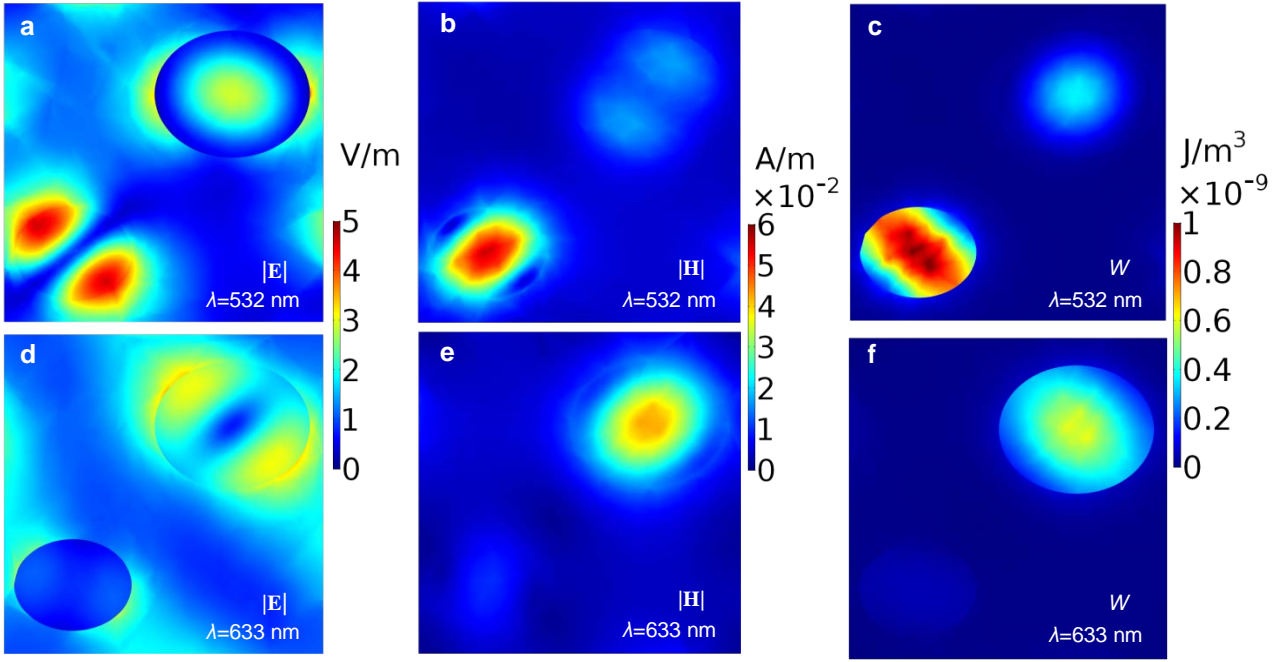

**Figure S1**. Near field distributions of electric, magnetic and power fields of two-size meta-atoms multiplexed in a supercell (*P*=320 nm). The sizes of elliptic nanocylinders are $l_x$=156 nm, $l_y$=128 nm and $l_x$=118 nm, $l_y$=92 nm, respectively, which manipulate the light at the wavelengths of 633 nm and 532 nm, respectively. (**a** - **c**) Distributions of electric (|**E**|), magnetic (|**H**|) and power (*W*) fields for the incident illumination at *λ*=532 nm. (**d** - **f**) Distributions of electric, magnetic and power fields for the incident illumination at *λ*=633 nm.

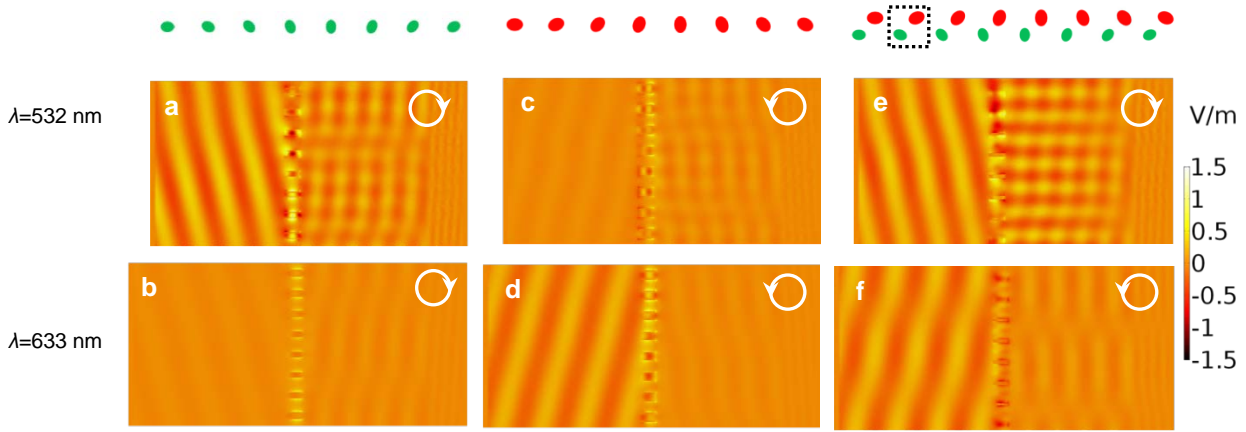

**Figure S2**. Electric field distributions of the periodic arrays consisting of single size and multiplexed two - size meta-atoms. (**a** - **b**) Electric field distributions of the array consisting of the meta-atom of $l_x$=118 nm, $l_y$=92 nm under the illumination of LCP light at $\lambda$=532 nm and 633 nm, respectively. (**c** - **d**) Electric field distributions of the array consisting of the meta-atom of $l_x$=156 nm, $l_y$=128 nm under the illumination of RCP light at $\lambda$=532 nm and 633 nm, respectively. (**e** - **f**) Electric field distributions of the array consisting of multiplexing two-size meta-atoms under the illumination of LCP green and RCP red light, respectively.

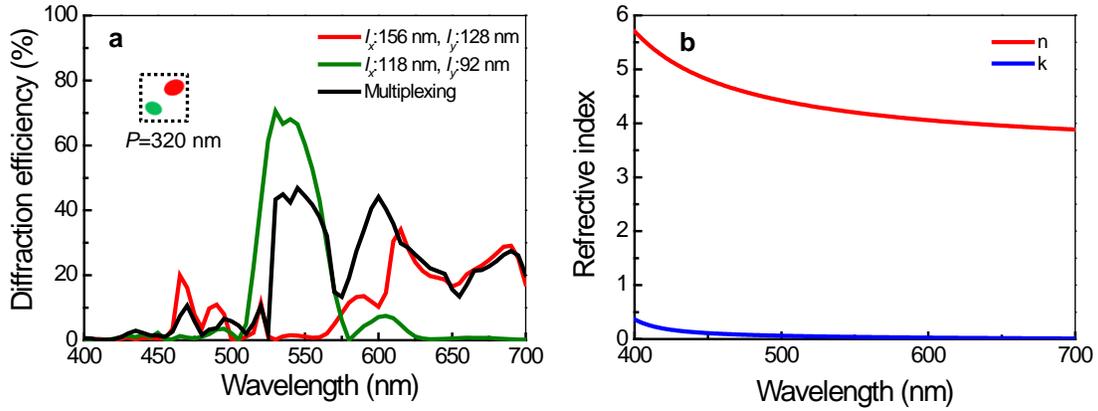

**Figure S3.** Simulated diffraction efficiencies of different elliptic nanocylinders. The diffraction efficiency is defined as the ratio of the reflected power of two-size meta-atoms with the period of $P$=320 nm for the circularly polarized light with the same helicity to the incident power. (**a**) Diffraction efficiencies of individual and multiplexed two-size meta-atoms in a super-cell. (**b**) Measured wavelength dependences of refractive index (real part $n$ and imaginary part $k$) of 230 nm thick Si layer.

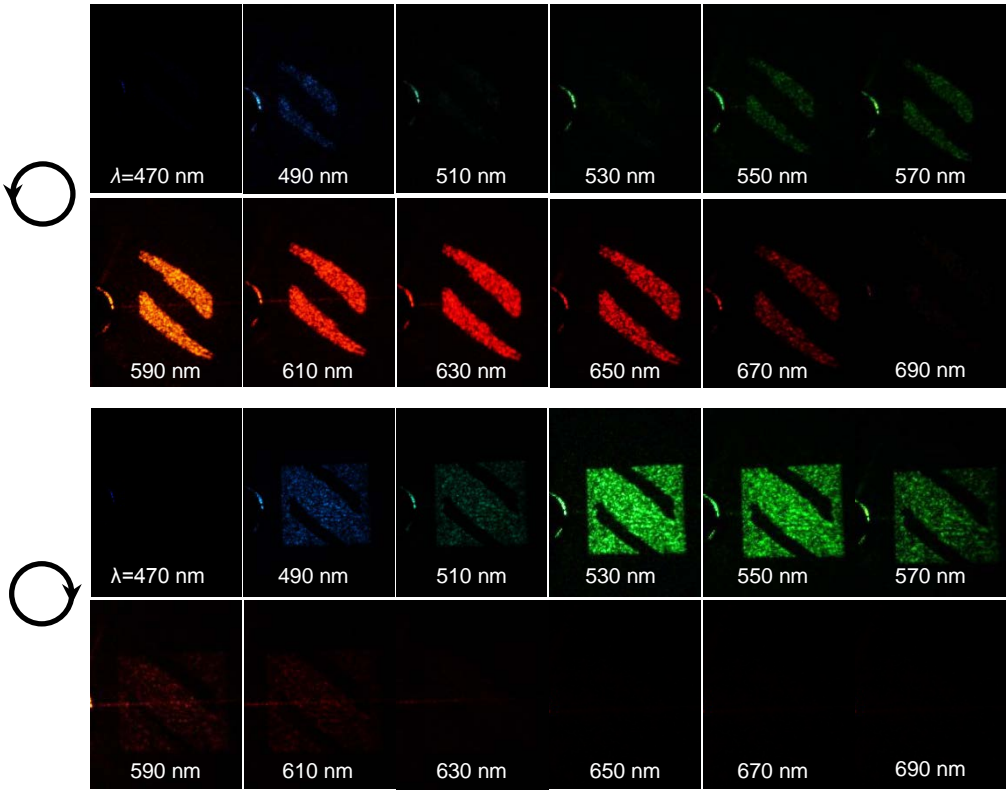

**Figure S4**. Wavelength dependence of encrypted information derived at the wavelengths ranging from 470 nm to 690 nm. The information is encoded in the RCP red and LCP green channels.

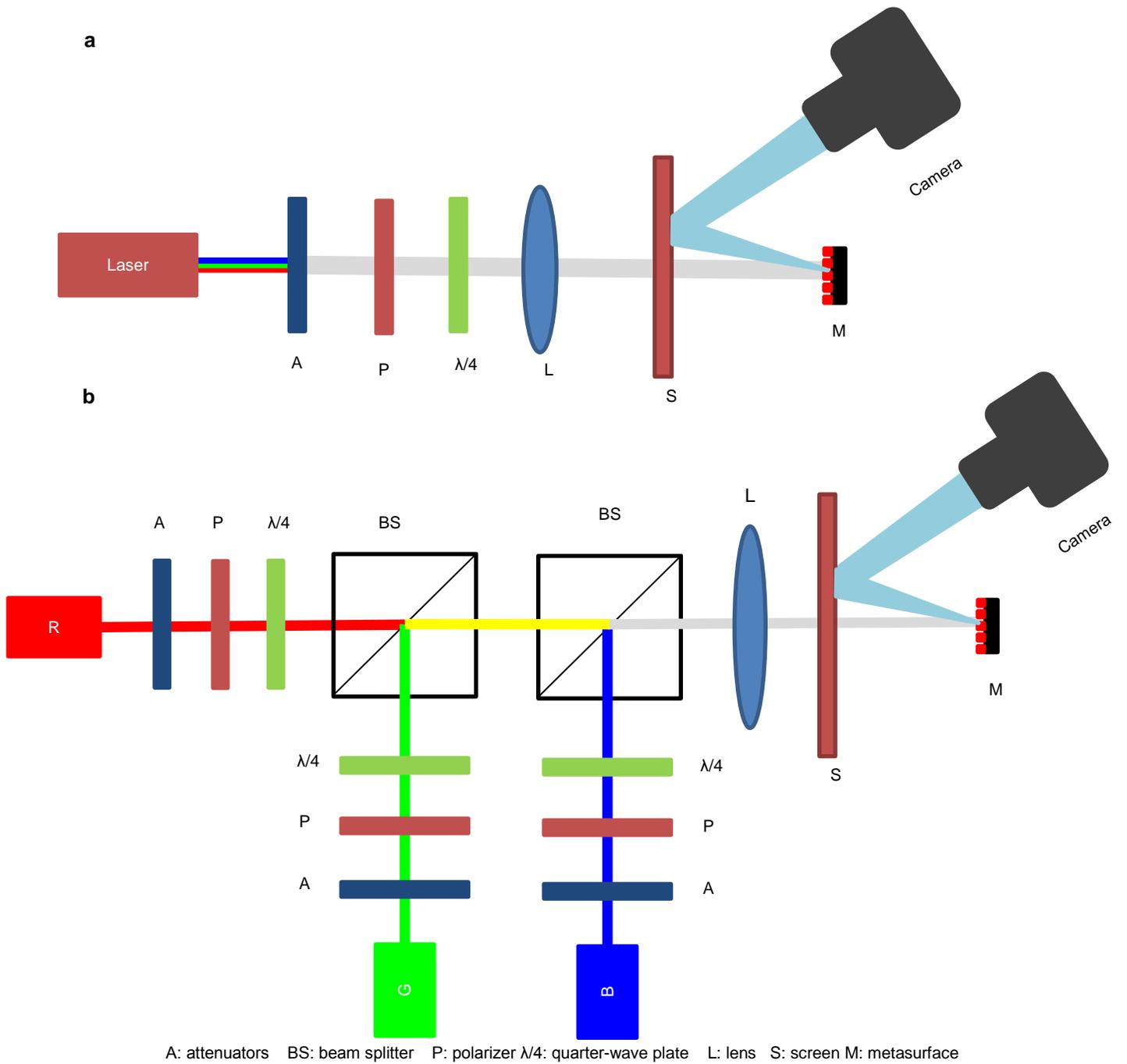

**Figure S5.** Setups for optical measurement. (**a**) Setup using a supercontinuum laser source. (**b**) Setup capable of individually controlling the polarization state.